\documentclass[12pt]{iopart}
\usepackage{bm}
\usepackage{harvard}
\usepackage{amssymb}
\usepackage
{hyperref}

\def\veps{\varepsilon}

\begin{document}
\title[Parity nonconservation in dielectronic recombination]
{Parity nonconservation in dielectronic recombination of multiply
charged ions}

\author{M G Kozlov$^1$,
G F Gribakin$^2$, and F J Currell$^2$}

\address{$^1$ Petersburg Nuclear Physics Institute,
Gatchina 188300, Russia}
\address{$^2$ School of Mathematics and Physics, Queen's University,
Belfast BT7 1NN, Northern Ireland, United Kingdom}
\ead{mgk@MF1309.spb.edu}

\date{\today }

\begin{abstract}
We discuss a parity nonconserving (PNC) asymmetry in the cross
section of dielectronic recombination of polarized electrons on
multiply charged ions with $Z \gtrsim 40$. This effect is strongly
enhanced for close doubly-excited states of opposite parity in the
intermediate compound ion. Such states are known for He-like ions.
However, these levels have large energy and large radiative widths
which hampers observation of the PNC asymmetry. We argue that
accidentally degenerate states of the more complex ions may be more
suitable for the corresponding experiment.
\end{abstract}




\section{Introduction}
\label{intro}

First discussion of parity nonconservation (PNC) effects in multiply
charged ions (MCI) appeared soon after formulation of electroweak
theory \cite{GL74}.
Since then many authors attempted to suggest realistic experiments
\cite{SSI89,KLN92,ZB97}. Recently PNC asymmetry in electron
scattering was calculated by \citeasnoun{MS05}. Here we focus on the
PNC asymmetry in dielectronic recombination (DR) on MCI. A
particular case of DR on H-like ions has been discussed in
\citeasnoun{GCK05}. PNC asymmetry there is strongly enhanced because of
the near-degeneracy of doubly-excited $2l2l'$ states of opposite
parity in He-like ions. Similar enhancement can take place for other
ions if there are close levels of opposite parity.
In this paper we use expressions for the PNC asymmetry derived by
\citeasnoun{GCK05} to analyze the optimal conditions for observation
of the PNC effects in DR and compare them with the H-like ions. We
show that if suitable levels in non-hydrogenic ions are found, the
statistical sensitivity of the PNC experiment can be improved by
several orders of magnitude.

\section{Feasibility of PNC experiment}
\label{basic}

In this section we briefly
outline the main results from \citeasnoun{GCK05}.
The Feasibility of measuring the PNC effect in DR depends on the
sensitivity requirements for an experimental apparatus to
observe the PNC asymmetry
\begin{equation}\label{d0}
     {\cal A} = \frac{\sigma^{+}-\sigma^{-}}{\sigma^{+}+\sigma^{-}},
\end{equation}
where $\sigma^\pm$ are the
recombination 
cross sections for electrons with
positive and negative helicity. One can consider two limiting cases:
the monoenergetic beam and the beam with wide energy distribution in
comparison to the energy difference between the
DR resonances of opposite parity $|E_1-E_2|$. For the most interesting
case of the very close resonances the typical experimental
conditions are likely to be closer to the latter case.
However, it is
easier to start with the  former, simpler case.

The number of counts in an
ideal 
experiment with a fully polarized electron beam
with positive helicity is given by:
\begin{equation}\label{d1}
  N_+= j_e N_i t \sigma^+ \equiv \mathcal{I}\sigma^+,
\end{equation}
where $j_e$ is the electron flux, $N_i$ is the number of target
ions, $t$ is the acquisition time. The number of counts for negative
helicity is $N_-=\mathcal{I}\sigma^-$.

For a beam with polarization $P$, to detect the PNC asymmetry, the
difference between the counts, $P |N_+ -N_-|$ should be greater than
statistical error, $\sqrt{N_+ +N_-}$, which gives:
\begin{equation}\label{d2}
  \mathcal{I} > \frac{\sigma^+ + \sigma^- + 2\sigma_b}{P^2(\sigma^+ - \sigma^-)^2},
\end{equation}
where $\sigma_b$ is the magnitude of any background occurring
through direct radiative recombination or as an experimental
artifact (e.g. detector dark counts). This is the effective cross
section to which the apparatus background would correspond. For the
rest of this analysis we consider the ideal limit, $P=1$, $\sigma_b
= 0$.

If the electron energy spread in the beam is greater than the
resonance spacing and widths, then the flux $j_e$ in \eref{d1}
should be replaced by the flux density $dj_e/d\varepsilon$. The
counts $N_{\pm}$ are obtained by integrating over the electron
energy and the effect can be detected if
\begin{equation}\label{d3}
  \mathcal{I}_{\rm av} > \int(\sigma^+ + \sigma^-) d \veps
  \left(\int (\sigma^+ - \sigma^-) d \veps\right)^{-2}.
\end{equation}
The first integral above is equal to $2(S_1+S_2)$, where
\begin{equation}\label{d4}
  S_i = \frac{\pi^2}{2p^2}\,
  \frac{\Gamma_i^{(r)}\Gamma_i^{(a)}}{\Gamma_i},
\end{equation}
is the strength of resonance $i$ (we use atomic units). In this
expression $p$ is the momentum of the incident electron,
$\Gamma^{(a)}_i$ and $\Gamma^{(r)}_i$ are autoionizing and radiative
widths of the resonances of opposite parity, $\Gamma_i=
\Gamma^{(a)}_i + \Gamma^{(r)}_i$ being their total widths.

The integral $\int (\sigma^+ - \sigma^-) d \veps$ in \eref{d3} can
be written as $2S_{1,2}^{\rm PNC}$, where
\begin{eqnarray}
  S_{1,2}^{\rm PNC} \equiv
  \int \sigma^{\rm PNC} d \veps
  &=& -\frac{\pi^2}{p^2}\,
     \frac{\sqrt{\Gamma_1^{(a)}\Gamma_2^{(a)}} h^{\rm PNC}
     \left({\Gamma_1^{(r)}}/{\Gamma_1}+{\Gamma_2^{(r)}}/{\Gamma_2}\right)}
     {(E_1-E_2)^2+\frac14(\Gamma_1 + \Gamma_2)^2}
\nonumber\\ &\times&
     \left[(E_1-E_2)\cos\delta_{1,2}
     +\case12(\Gamma_1 +\Gamma_2)
     \sin\delta_{1,2} \right],
\label{d5}
\end{eqnarray}
is the PNC strength of the two resonances. Here $h^{\rm PNC}$ is the
PNC matrix element which mixes the resonances, $E_i$ are their
energies, and $\delta_{1,2}$ is the relative Coulomb phase between
two channels
(see \citeasnoun{GCK05} for details). 
\Eref{d3} now reads:
\begin{equation}\label{d6}
  \mathcal{I}_{\rm av} > \case12(S_1+S_2)/\left(S_{1,2}^{\rm PNC}\right)^2.
\end{equation}

We see that for the limiting case of wide energy distribution in the
beam the feasibility of the PNC experiment depends on the function:
\begin{equation} \label{d8}
F_{\rm av} = \int(\sigma^+ + \sigma^-) d \veps \left[\int (\sigma^+
- \sigma^-) d \veps\right]^{-2}\!\!.
\end{equation}
Optimal experimental conditions correspond to the minimum of the
function $F_{\rm av}$. In the next section we examine how this
function depends on the parameters of the resonances.

\section{Optimization}
\label{opt}

Substituting equations~\eref{d4} and \eref{d5} in \eref{d8} we obtain an
explicit expression for $F_{\rm av}$:
\begin{equation} \label{o1}
\fl
  F_{\rm av} =
  \frac{p^2}{4\pi^2}
  \frac{{\Gamma^{(r)}_1 \Gamma^{(a)}_1}/{\Gamma_1}
        +{\Gamma^{(r)}_2 \Gamma^{(a)}_2}/{\Gamma_2}}
  {\Gamma^{(a)}_1 \Gamma^{(a)}_2 \left(h^{\rm PNC}\right)^2
  \left({\Gamma^{(r)}_1}/{\Gamma_1}
        +{\Gamma^{(r)}_2}/{\Gamma_2}\right)^2}
  \left(\frac{\Delta^2+\Gamma^2}
             {\Delta\cos\delta_{1,2}+\Gamma\sin\delta_{1,2}}
  \right)^2,
\end{equation}
where $\Delta \equiv E_1-E_2$ and $\Gamma\equiv
\case{1}{2}(\Gamma_1+\Gamma_2)$. Function $F_{\rm av}$ has
dimension $\mathrm{area}^{-1}\times \mathrm{energy}^{-1}$. It is
convenient
to write it as a product, $F_{\rm av} = F_1 \times F_2 \times F_3$,
where $F_1$ is a dimensional factor, and $F_3$ depends only on $\Delta$
and $\Gamma$, but not on $\Gamma^{(a)}_i$ or $\Gamma^{(r)}_i$:
\numparts
\begin{eqnarray} \label{o3a}
  F_1 &=
  {p^2\Gamma}\left(2\pi h^{\rm PNC}\right)^{-2}\,,
\\  \label{o3b}
  F_2 &=
  \frac{\Gamma \left({\Gamma^{(r)}_1 \Gamma^{(a)}_1}/{\Gamma_1}
        +{\Gamma^{(r)}_2 \Gamma^{(a)}_2}/{\Gamma_2}\right)}
  {\Gamma^{(a)}_1 \Gamma^{(a)}_2
  \left({\Gamma^{(r)}_1}/{\Gamma_1}
        +{\Gamma^{(r)}_2}/{\Gamma_2}\right)^2}\,,
\\  \label{o3c}
  F_3 &=
  {\left(\Delta^2+\Gamma^2\right)^2}
             \Gamma^{-2}\left(\Delta\cos\delta_{1,2}
              +\Gamma\sin\delta_{1,2}\right)^{-2}\,.
\end{eqnarray} \label{o3}
\endnumparts
Since the 
three factors
above depend on different parameters, it is convenient to analyze
them separately. The dimensional factor \eref{o3a} depends
on the PNC matrix element,
the average DR width $\Gamma $ 
and the electron momentum $p$,
which is defined by the energy difference between the ground state $E_0$
of the target ion and the autoionizing resonances, $p^2/2
= E_{1,2}-E_0$. Thus, the optimal conditions correspond not only to
the close levels of opposite parity, but also to the lowest possible
energy of the beam. For H-like ions $p^2\approx Z^2/2$ and $|h^{\rm
PNC}|\approx 1.1 \times 10^{-18} Z^5$
\cite{GCK05}. 
Thus, the overall scaling of
the first factor is:
\begin{equation} \label{o4}
  F_1 (\mbox{H-like})
  \approx 1.0 \times 10^{34} Z^{-8}\Gamma\,.
\end{equation}
This expression shows that it
would be 
much easier to observe PNC effects
in heavy ions. Below we will focus on $Z > 30$, where $\Gamma$ is
dominated by radiative widths $\Gamma_{1,2}^{(r)}$ and grows rapidly
with $Z$:
\begin{equation} \label{o4a}
\Gamma \approx 0.76 \times 10^{-8} Z^4\,,
\end{equation}
which yields the following 
estimate:
\begin{equation} \label{o4b}
  F_1 (\mbox{H-like})
  \approx 0.76 \times 10^{34} Z^{-4}\,.
\end{equation}

We see that because of the growth of $p^2$ and $\Gamma$ with $Z$,
the overall scaling of $F_1$ is much weaker than one might expect
from the scaling of the matrix element $h^{\rm PNC}$ alone. For
non-hydrogenic MCI this matrix element is of the same order of
magnitude. Therefore, if one could find low-lying autoionizing
levels of opposite parity, it would be possible to reduce both $p$
and $\Gamma$ in \Eref{o3a}.

The two remaining factors, $F_2$ and $F_3$, are dimensionless. Factor
$F_3$ depends on the relation between $\Gamma$ and $\Delta$: $F_3
\approx \sin^{-2}\!\delta_{1,2}$ for $\Gamma\gg \Delta$, and
$F_3\approx (\Delta /\Gamma )^2\cos^{-2}\!\delta_{1,2}$ 
for $\Gamma\ll \Delta$.
Optimization of this factor requires that $\Gamma > \Delta$. For
H-like ions this condition is met for $Z\gtrsim 40$, where
$F_3$ is practically optimal and of the order of unity.

Finally, we turn to the factor $F_2$ which depends on the relative
size of the autoionizing and radiative widths. Consider first the
``symmetric'' case $\Gamma^{(r)}_i=\Gamma^{(a)}_i$. Then
$F_2=2\Gamma^2/(\Gamma_1\Gamma_2)$ and the minimum is reached for
$\Gamma_1=\Gamma_2$:
\begin{equation}\label{o5}
\left.\min F_2\right |_{\Gamma^{(r)}_i=\Gamma^{(a)}_i}=2.
\end{equation}
In the ``asymmetric'' case $\Gamma^{(r)}_i\ne\Gamma^{(a)}_i$ the
denominator becomes small if either both radiative widths, or both
autoionizing widths become small. Therefore, optimal conditions
correspond to the case when one radiative widths is large and
another is small. Let us assume that $\Gamma^{(r)}_1 \ll
\Gamma^{(a)}_1$ and $\Gamma^{(r)}_2 \gg \Gamma^{(a)}_2$. That gives
\begin{equation}\label{o6}
  F_2 = \frac{\Gamma}{\Gamma_1}\,
  \frac{\Gamma^{(r)}_1+\Gamma^{(a)}_2}{\Gamma^{(a)}_2}\,.
\end{equation}
Now the minimum is reached for $\Gamma_1\gg\Gamma_2$ and
$\Gamma^{(r)}_1\ll\Gamma^{(r)}_2$:
\begin{equation}\label{o7}
\left.\min
F_2\right|_{\Gamma^{(r)}_1\ll\Gamma^{(a)}_1}=\case{1}{2}\,.
\end{equation}

We see that it is beneficial
if the width of the narrower resonance is dominated by the radiative
decay, while that of the broader one~---~by autoionization. However,
the gain in comparison to the symmetric case is only a factor of 4.
It is important though to have at least one radiative and one
autoionizing width of the order of the total width $\Gamma$. In
H-like ions the autoionizing widths of both resonances are
smaller than $\Gamma ^{(r)}_i$ for
$Z > 30$. Using expressions for the widths from \citeasnoun {GCK05}
we can estimate the $F_2$ factor for H-like ions:
\begin{equation}\label{o8}
\left. F_2(\mbox{H-like})\right|_{Z> 30}\approx 2.3 \times 10^{-6}
Z^4\,.
\end{equation}
For H-like ions with $40\lesssim Z\lesssim 60$ the function $F_2$ is
1~--~2 orders of magnitude larger than minimal value \eref{o7}.
Taking into account equation \eref{o4b} and the fact that for the H-like ions
with $Z\gtrsim 40$ the factor $F_3$ is of the order of unity we get
the estimate:
\begin{equation}\label{o9}
\left. F_\mathrm{av}(\mbox{H-like})\right|_{Z\gtrsim 40} \approx 1.7
\times 10^{20}\,.
\end{equation}

Thus, in spite of the strong dependence of the PNC interaction on
$Z$, the feasibility function for the H-like ions is practically
independent of $Z$.
The main reason for this is the rapid growth of the radiative widths of
the resonances, $\Gamma^{(r)} \propto
\omega^3 |E1|^2$. For H-like ions the radiative transition frequency
is $\omega\approx \case38 Z^2$, and the transition amplitude $E1$ decreases
as $Z^{-1}$, so that $\Gamma^{(r)} \propto Z^4$.

For non-hydrogenic ions the transition frequencies are typically much
smaller, while the $E1$ amplitudes are of the same order of magnitude.
Therefore we can expect smaller values of the functions $F_1$ and
$F_2$. In addition, for non-hydrogenic ions the autoionizing resonances
typically lie at lower energies. This means that the electron momentum $p$
in equation \eref{o3a} is smaller. Of course, it is still
necessary to find closely-spaced resonances of opposite parity with
$\Delta \sim \Gamma$,
to keep the function $F_3$ small. If such resonances are found for ions
with $Z \gtrsim 40$, one can expect a noticeable improvement of the
experimental sensitivity to the PNC asymmetry in comparison with similar
H-like ions.

\section{Conclusions}

We see that the feasibility of observing PNC asymmetry in DR on H-like
ions is increased by the proximity of the autoionizing resonances of
opposite parity and decreased by the large energy and radiation
widths of these resonances. For non-hydrogenic ions the autoionizing
resonances usually lie at lower energies and have smaller radiative
widths, but are further apart. On the other hand, the number of such
ions is huge and there can be accidental degeneracies between levels
of opposite parity. Level widths grow rapidly with $Z$, increasing
the probability to find pairs of levels with $\Delta \sim \Gamma$
for heavier ions where PNC effects are larger. Finding such close
levels for ions with $Z\gtrsim 40$ can significantly simplify
observation of the PNC asymmetry in comparison to H-like ions.

\vspace{5mm}\noindent
This work was partly supported by the Russian
Foundation for Basic Research, grant No. 05-02-16914.
\section*{References}


\end{document}